\documentclass[sigconf]{acmart}

\usepackage{url}
\usepackage{hyperref}

\AtBeginDocument{%
  }

\setcopyright{acmcopyright}
\copyrightyear{2023}
\acmYear{2023}
\acmDOI{}

\acmConference[HCPS '23]{HCPS '23 Workshop on Humans in Cyber-Physical Systems}{May 09--12,
  2023}{San Antonio, TX}
\acmPrice{15.00}
\acmISBN{}

\begin{document}

\title{Refining Human-Centered Autonomy Using Side Information}

\author{Adam J. Thorpe}
\email{ajthor@unm.edu}
\orcid{0000-0001-7120-0913}
\affiliation{%
  \institution{University of New Mexico}
  \city{Albuquerque}
  \state{New Mexico}
  \country{USA}
}



\begin{abstract}
Data-driven algorithms for human-centered autonomy use observed data to compute models of human behavior in order to ensure safety, correctness, and to avoid potential errors that arise at runtime. However, such algorithms often neglect useful a priori knowledge, known as side information, that can improve the quality of data-driven models.
We identify several key challenges in human-centered autonomy, and identify possible approaches to incorporate side information in data-driven models of human behavior.
\end{abstract}




\maketitle

\section{Introduction}

As autonomous technologies are developed and deployed in practical, real-world environments, they must necessarily take humans into account during the design phase. 
However, humans in the loop pose a significant, enduring challenge for autonomous systems--in particular since humans are notoriously difficult to model and predict, and the lack of simple mathematical models hinders the use of traditional, well-established control techniques.
In addition, humans are by nature inconsistent, heterogeneous, and non-stationary, meaning any single, static mathematical model is likely insufficient to capture the wide array of potential observable human behaviors. 
As a result, it is likely impossible to precompute an autonomous control algorithm that will demonstrate good performance in all contexts or for all humans.
%
This presents a need for algorithms that are adaptive to new, unseen scenarios, responsive to user preferences, and that can be computed using observed data that is available at runtime--possibly from a single, evolving trajectory. 

Data-driven methods for autonomy present one possible approach to account for humans in the loop. However, these methods bring new challenges. In particular, data-driven autonomy is naturally data-dependent, meaning it is reliant upon observed phenomena captured by the training data to perform inference.
As such, it can be fragile, and is often expected to generalize far outside the data regime, which is unreasonable in many cases. 
This reflects a fundamental challenge in learning theory. 
By construction, learning-based or data-driven algorithms for autonomy are often designed to take an unbiased view of the data in an attempt to minimize prior assumptions made on the learned function.
However, such an unbiased view neglects prior knowledge or structure which may be useful in modeling human decision-making processes. 

We propose that prior knowledge of human behavior and decision-making, sometimes known as side information, can offer useful insights to enhance human-centered autonomous control algorithms.
For instance, incorporating knowledge of psychological, physiological, or environmental factors that affect decision-making (which are well studied in the context of human factors). 
Nevertheless, applying these behavioral insights to existing data-driven autonomous control algorithms poses a significant challenge. 
This motivates the need for new methods that can incorporate such insights into autonomous control design. 

\subsection*{Related Work}


Existing algorithms for human-centered autonomous control often rely upon simple descriptions of human behavior to enable tractability, or use statistical learning tools which yield a black-box representation that can be difficult to interpret or analyze \cite{rudenko2020human}. 
Such models are either too general or too highly tailored to specific contexts or scenarios \cite{179417}, and often rely upon simplifying assumptions of Gaussianity or linearity, which can lead to a model mismatch or unpredictable behaviors. 
%
Further, existing models generally ignore side information or data available at runtime. 
%

Recently, work in the area of adapting data-driven algorithms to incorporate side information or bias have been explored in the context of stochastic control \cite{https://doi.org/10.48550/arxiv.2301.03565, pmlr-v168-djeumou22a, pmlr-v120-ahmadi20a}.
Further, data-driven methods that can incorporate side information and learn using a single trajectory have been explored in \cite{9930630, 9483367, pmlr-v168-djeumou22b}.
However, these methods often consider side information in the form of knowledge of the system dynamics, algebraic constraints, or mathematical properties, and it is unclear whether these existing methods can be adapted to take into account the types of information relevant to human decision-making processes. 

\section{Challenges}

Here, we present some common challenges that apply across problem domains. 

\paragraph{Dealing With Human Data} 

Human data is messy and highly variable, even in highly controlled environments. 
Variations caused by psychophysiological, environmental, or behavioral factors can have a dramatic impact on human decision-making and observed human behaviors.
Moreover, these factors can change rapidly, and are often difficult to measure or predict, leading to a high amount of uncertainty. 
In particular, human data is stochastic, and can depend on myriad factors which are opaque to an autonomous controller. 
%
Thus, a core challenge in human-centered autonomy is designing algorithms which are able to quickly adapt as data becomes available. 
In other words, such autonomous control algorithms must be adaptive to data available to the system at runtime. 
Furthermore, data-driven algorithms typically require large amounts of data collected upfront to synthesize a learned model. As such, adapting existing methods to low-data regimes, and taking real-time data into account remain challenging problems. 

\paragraph{Extending Techniques to Incorporate Side Information}

Data-driven algorithms are typically designed to avoid implicit bias in order to promote generality. 
While an unbiased approach can be useful to identify unknown relationships in data, bias can be useful in the context of dynamical systems since it can impart known structure on the data-driven model \cite{https://doi.org/10.48550/arxiv.2301.03565}. 
For instance, in the context of human behavioral models, such bias (i.e.\ side information) could encode prior knowledge of human decision-making processes in the data-driven autonomous controller. 
In addition, imparting bias in the learned solution can have the added benefit of reducing the overt dependence on the data and can account for behaviors which are difficult to discern solely from the data. 
%
Thus, a central challenge is designing data-driven algorithms that can incorporate side information, and that can accommodate types of side information that are not easily encoded into mathematical forms. 




\section{Incorporating Side Information}

Incorporating side information and adapting algorithms to account for human data 
can overcome some of the limitations present in existing behavioral models. 
In this section, we identify possible approaches to incorporate side information in data-driven autonomy. 

We focus primarily on learning-based models, that use data collected from observations of the system evolution to compute a model of human decision-making.
Learning-based models are highly flexible, meaning they can easily be adapted to different problem domains, and are primarily useful in identifying unknown functional relationships in observed data.
Broadly speaking, the goal of learning-based algorithms is to `fit' a learned function to observed data. 
This is useful in the case of human-centered autonomy, where the underlying process of human decision-making is complex and difficult to characterize. 
Of particular interest to the case of human data are learning-based algorithms which learn statistical quantities, such as \emph{distributions} over the data.

In mathematical terms, learning problems generally solve an underlying optimization problem, e.g.\ least-squares, to find a function that minimizes the empirical error over the observed data. 
By design, learning algorithms generally avoid implicit bias--a central tenet being that the simplest, or least complex solution is often the most generalizable. 
However, such bias can be useful, since it can encode prior knowledge of the solution.
For example, suppose we have (potentially imperfect) model information (such as approximate knowledge of human decision-making processes), and are also given data taken through system observations. 
We can encode such information in the learning problem by specifying constraints on the underlying optimization problem or by penalizing solutions which differ significantly from the known, approximate model--a process which we describe as, ``shifting the regression baseline.''
Incorporating imperfect information as bias in the learning problem can lead to a more accurate solution, since it considers \emph{both} the data and the side information in the learned behavior model.
In practical terms, this means we can learn the difference between the imperfect model and what is shown by the data. 

In addition, statistical learning-based algorithms typically impose structure on the learned function in order to ensure the problem is well-posed. 
Often, this means the solution to the learning problem is an element in a high-dimensional function space, e.g.\ a Hilbert space of functions. 
Such structure has the added benefit of providing useful mathematical tools for analysis, such as distance metrics which can determine, e.g.\ how far a human differs from the `mean'.
For example, suppose we learn a probabilistic behavioral model for two different human drivers in a shared autonomy setting, and we wish to identify how `close' the behavior of the drivers is. 
Because human data is often limited and highly stochastic, identifying the overall similarity between driver behavior presents a challenging problem. 
However, using a probability metric defined in the function space, we can easily determine the similarity between the two empirical, learned behavioral distribution models. 
For instance, in the case of Hilbert spaces of functions, this metric is known as the maximum mean discrepancy (MMD), and has proven exceptionally useful in statistical hypothesis testing \cite{JMLR:v13:gretton12a}.
In the case of learning distributions of human behavior, such tools can provide a useful means for the analysis and assessment of driver behavior. 

Lastly, human data is often limited, and data-driven algorithms typically require significant amounts of data. 
In order to be amenable to human data, learning-based algorithms will need to account for low-data regimes.
One possibility is to focus on online statistical learning techniques or recursive algorithms that can incorporate data collected online. 
As shown in \cite{https://doi.org/10.48550/arxiv.2301.03565},  incorporating side information side information has the added benefit of requiring less data in the learning problem.

\section{Conclusions}

Despite the recent advances in incorporating side information in data-driven autonomy, several key challenges remain. 
Here, we identify two such challenges in data-driven human autonomy: adapting existing techniques to the types of data in human-in-the-loop systems, and finding ways to incorporate side information in data-driven models of human behavior.
In addition, we have identified possible approaches to adapting learning-based algorithms to tackle such challenges. By adapting data-driven algorithms to account for side information, we can alleviate the overt dependence on data and encode prior knowledge of human decision-making into autonomous controllers. 
Of particular interest in this area are methods to parameterize learned models to be sensitive to changes in operating contexts, user preferences, or behavioral factors, and is an area of current research.

\bibliographystyle{ACM-Reference-Format}
\bibliography{bibliography.bib}

\end{document}